\documentclass{PoS}

\title{On the coverage of the pMSSM by simplified model results}

\ShortTitle{On the coverage of the pMSSM by simplified model results}

\author{Ursula Laa\thanks{For the SModelS collaboration}\\
        Laboratoire de Physique Subatomique et de Cosmologie,
Universit\'e Grenoble-Alpes,\\ \ \  CNRS/IN2P3, 53 Avenue des Martyrs, F-38026 Grenoble, France\\
and LAPTh, Universit\'e Savoie Mont Blanc, CNRS, B.P.110 Annecy-le-Vieux,\\ \ \  F-74941 Annecy Cedex, France\\
        E-mail: \email{ursula.laa@lpsc.in2p3.fr}}

\abstract{
ATLAS and CMS have been performing a large number of searches for physics beyond the Standard Model (BSM).
In particular results of supersymmetry (SUSY) searches are typically interpreted in the context of simplified models.
While mass limits obtained in this manner are highly model dependent, cross section upper limits
(or efficiency maps) may be used to obtain constraints on generic BSM scenarios without any further
event and detector simulation.
This procedure has been automatised in the public tool SModelS, which first decomposes the generic
scenario into its simplified model components which can then be tested directly against the experimental constraints in the
SModelS database.
We briefly introduce SModelS and then discuss how the coverage by simplified model results compares to
what can be obtained in a full simulation study for the example of the 19 parameter phenomenological
MSSM (pMSSM).
Considering all parameter points that ATLAS has tested in a comprehensive study~\cite{Aad:2015baa},
we find that about 60\% of the points excluded by ATLAS can be excluded from simplified model
constraints.
This fraction could be improved by considering currently `missing' simplified model topologies.
We show that in particular topologies with asymmetric branches are often important, while long cascade
decays with more than one intermediate particle are less relevant.
A particularly interesting asymmetric branch topology is a 3 jet + $E_T^{\mathrm{miss}}$ signature arising
from gluino-squark production.
}

\FullConference{The European Physical Society Conference on High Energy Physics\\
		5-12 July, 2017\\
		Venice}

\begin{document}

\section{Introduction}
Simplified model spectra (SMS) have become one of the standard methods to interpret BSM
searches at the LHC.
Simplified model results can be used to constrain generic models, providing a fast
test against a large number of existing searches.
It is generally assumed that while simplified model based reinterpretation is much faster than
event and detector simulation based reinterpretation, it is significantly less accurate and/or
general, typically giving a conservative estimate of the true limits.
Here we aim to quantify the difference in exclusion potential for the example of the 19-parameter pMSSM, based
on the ATLAS study~\cite{Aad:2015baa} in which the points from an extensive
pMSSM scan were tested against the constraints from 22 ATLAS searches from LHC Run 1.
ATLAS made the SLHA spectra of the whole scan public on HEPDATA~\cite{hepdata}
together with information about which point is excluded by which analyses.
This is extremely useful information, which we here use to test the constraining power of
SMS results by means of SModelS~\cite{Kraml:2013mwa,Ambrogi:2017neo}.

SModelS is an automatised tool for interpreting simplified model results from the LHC.
It decomposes collider signatures of new physics featuring a $\mathbb{Z}_2$-like symmetry into
simplified model topologies, using a generic procedure where each SMS is defined by the vertex
structure and the Standard Model (SM) final state particles;
BSM particles are described only by their masses, production cross sections and branching ratios.
The weights of the various topologies, computed as production cross section times branching ratios,
are then compared against a large database of experimental constraints.
This procedure takes advantage of the large number of simplified models already constrained by
official ATLAS and CMS results and does not require Monte Carlo event simulation,
thus providing a fast way of confronting a full BSM model with the LHC constraints.
Furthermore, `missing' topologies, which are not covered by any of the experimental constraints,
are identified and provided as an output of SModelS.

\section{Setup of the analysis}
In~\cite{Aad:2015baa} ATLAS has analysed in total more than 310k pMSSM parameter
points with SUSY masses below 4 TeV and a neutralino as the lightest SUSY particle (LSP).
These points satisfy constraints from previous collider searches, flavor and electroweak (EW)
precision measurements, cold dark matter relic density and direct dark matter searches.
In addition, the mass of the light Higgs boson was required to be between 124 and 128 GeV.
These points were classified into three sets according to the nature of the LSP:
bino-like (103410 points), wino-like (80233 points) and higgsino-like (126684 points).
About 40\% of all these points were excluded by at least one of the 22 ATLAS Run 1 searches
considered in the analysis.

These points excluded by ATLAS are the centre of interest of our study.
The reason is that we want to compare the exclusion coverage obtained using
SMS results only to what is obtained in a full simulation study.
We restrict our analysis to the sets with bino-like (42039 points) or higgsino-like (48703 points) LSP, neglecting points with a wino-like LSP,
as most of them lead to a displaced vertex signature
which cannot be studied with the current version of SModelS.
We further remove points from the bino- and higgsino-like LSP data sets if they contain any long lived
sparticles---this concerns however only a small number of points.
Likewise, points which ATLAS found to be excluded only by heavy Higgs searches are also not considered here, because no corresponding
searches are included in the SModelS database.
This selection leaves us with 38,575 parameter points with a bino-like LSP and
45,594 parameter points with a higgsino-like LSP to be tested.
We use version 1.1.1 of SModelS, which works with upper limit (UL) and efficiency map (EM) type
results~\cite{Ambrogi:2017neo}.
The official ATLAS and CMS Run~1 results were augmented with several `home-grown' efficiency maps in the
v1.1.1 database and we further extend this database with Fastlim~\cite{Papucci:2014rja} efficiency maps.
For details about the used results database as well as details on the cross section computation we refer
the reader to~\cite{Ambrogi:2017neo,Ambrogi:2017lov}

\section{Results}
The results are summarized in Table~\ref{tab:smodpmssmsum}, where we list the total number of points
studied, the number of points that can be excluded by SModelS when using only the upper limit
results in the database, and the number of points that can be excluded when using the full 8~TeV database,
that is including efficiency map results.
We see that in particular the coverage of bino-like LSP scenarios can be improved by using efficiency maps.
Concretely the coverage improves from 44\% (UL results only) to 55\% (full database).
Similarly the coverage for the higgsino-like LSP scenarios is increased from 55\% to 63\%.
In the following we will focus only on bino-like LSP scenarios with light gluinos,
an extensive discussion including also the higgsino-like LSP scenarios can be found
in~\cite{Ambrogi:2017lov}.

\begin{figure}[t!]\centering
\includegraphics[width=0.45\textwidth]{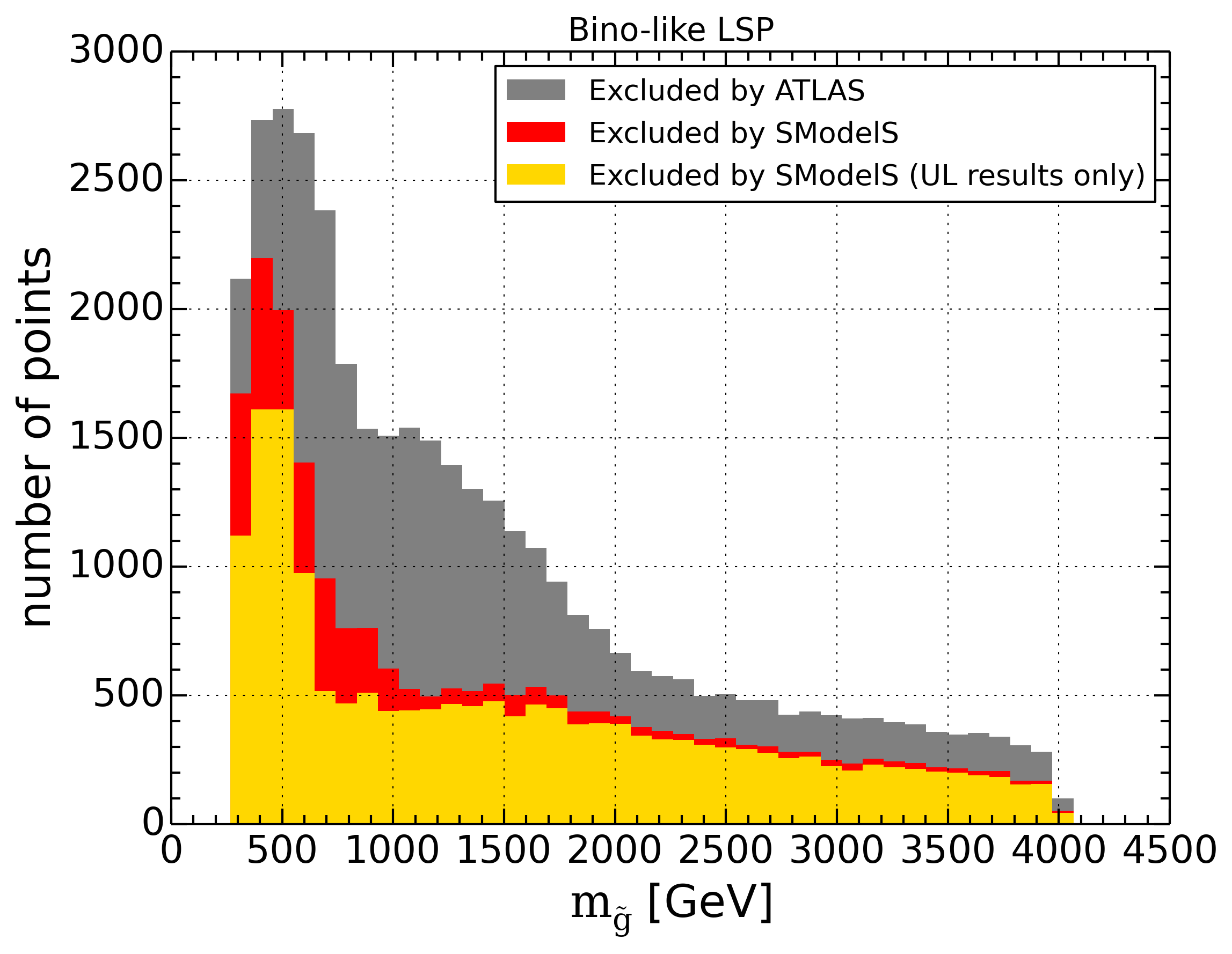}\hfill
\includegraphics[width=0.53\textwidth]{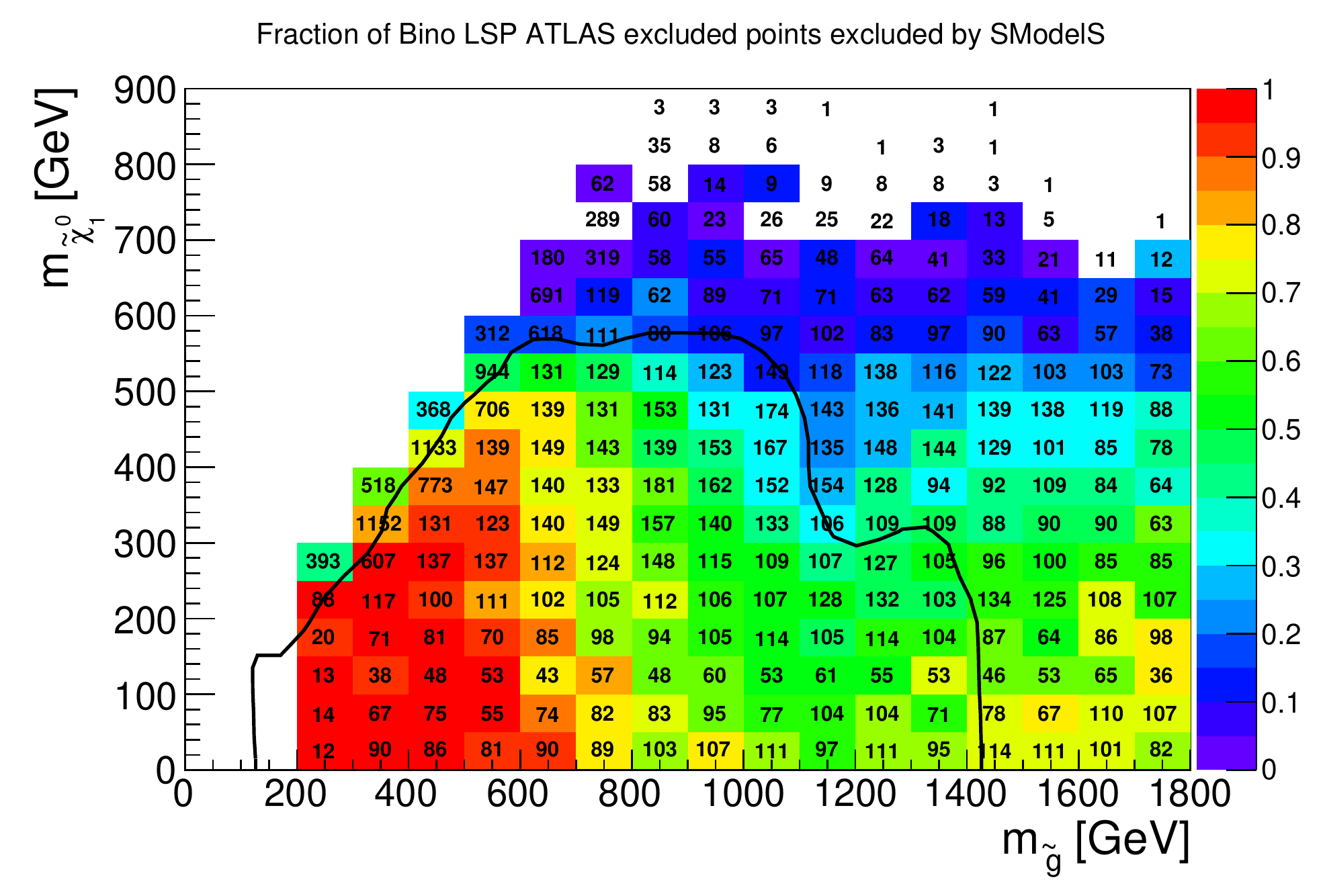}
\caption{Left: Number of points excluded by SModelS  when using the full 8~TeV database (in red) and when
using only UL results (in yellow),
as a function of the gluino mass.
For reference the total number of ATLAS-excluded points is also shown (in grey).
Right: Coverage in the gluino vs.\ neutralino mass plane, for gluino masses up to $2$~TeV.
The color code indicates the fraction of points excluded by SModelS, the
text gives the total number of points tested in each bin.
For comparison the exclusion line obtained in~\cite{Aad:2014wea} for a simplified model
where gluino pair production is followed by the direct decay
$\tilde{g}\rightarrow q q \tilde{\chi}^0_1$ is drawn in black.}
\label{fig:pmssmGlu1d}
\end{figure}

Indeed the improved coverage is largely due to additional constraints on scenarios with light gluinos
when including efficiency map results,
as illustrated in Figure~\ref{fig:pmssmGlu1d} (left).
The main reason is that efficiency maps allow us to combine the signal for all topologies contributing
to the same signal region before comparing
against an overall cross section limit.
Moreover, some asymmetric decay branches are included in the EM-type results but not in
the UL-type results in the database.
Therefore, while UL results often constrain only a fraction of the total gluino
production (determined by the gluino decay branching ratios),
this can be improved when using EM results.

\begin{table}
\centering
\begin{tabular}{c | c | c}
 & Bino-like LSP & Higgsino-like LSP\\
\hline
Total number of points & 38575 & 45594 \\
\hline
Number of points excluded -- UL results only & 16957 & 25005 \\
\hline
Number of points excuded -- full database & 21171 & 28659
\end{tabular}
\caption{
Summary of results, listing the number of ATLAS-excluded pMSSM points tested in this study,
the number of points excluded by SModelS when
using UL-type results only, and the number of points excluded when using the full 8~TeV
database including EM-type results.}
\label{tab:smodpmssmsum}
\end{table}

Note, however, that there are still many points with light gluinos which cannot be excluded by the
SMS results in the SModelS database.
To understand this better we show in Figure~\ref{fig:pmssmGlu1d} (right) the coverage in the
gluino vs.\ neutralino mass plane.
For comparison the exclusion line obtained in~\cite{Aad:2014wea} for a simplified model where pair-produced gluinos decay exclusively as
$\tilde{g}\rightarrow q q \tilde{\chi}^0_1$ is also drawn.
We see that light gluinos escape SMS limits especially in the compressed region where monojet type searches become important.
This is in agreement with the example exclusion line.
Moreover, while the coverage is good for very light gluinos up to about $600$~GeV,
it drops for intermediate gluino masses around $1$~TeV and higher.
Concretely the coverage is 80\% when considering only points with light gluinos
($m_{\tilde{g}} < 600$~GeV), but drops
to 60\% when considering all points with $m_{\tilde{g}} < 1400$~GeV.

\begin{figure}[t!]\centering
\includegraphics[width=0.49\textwidth]{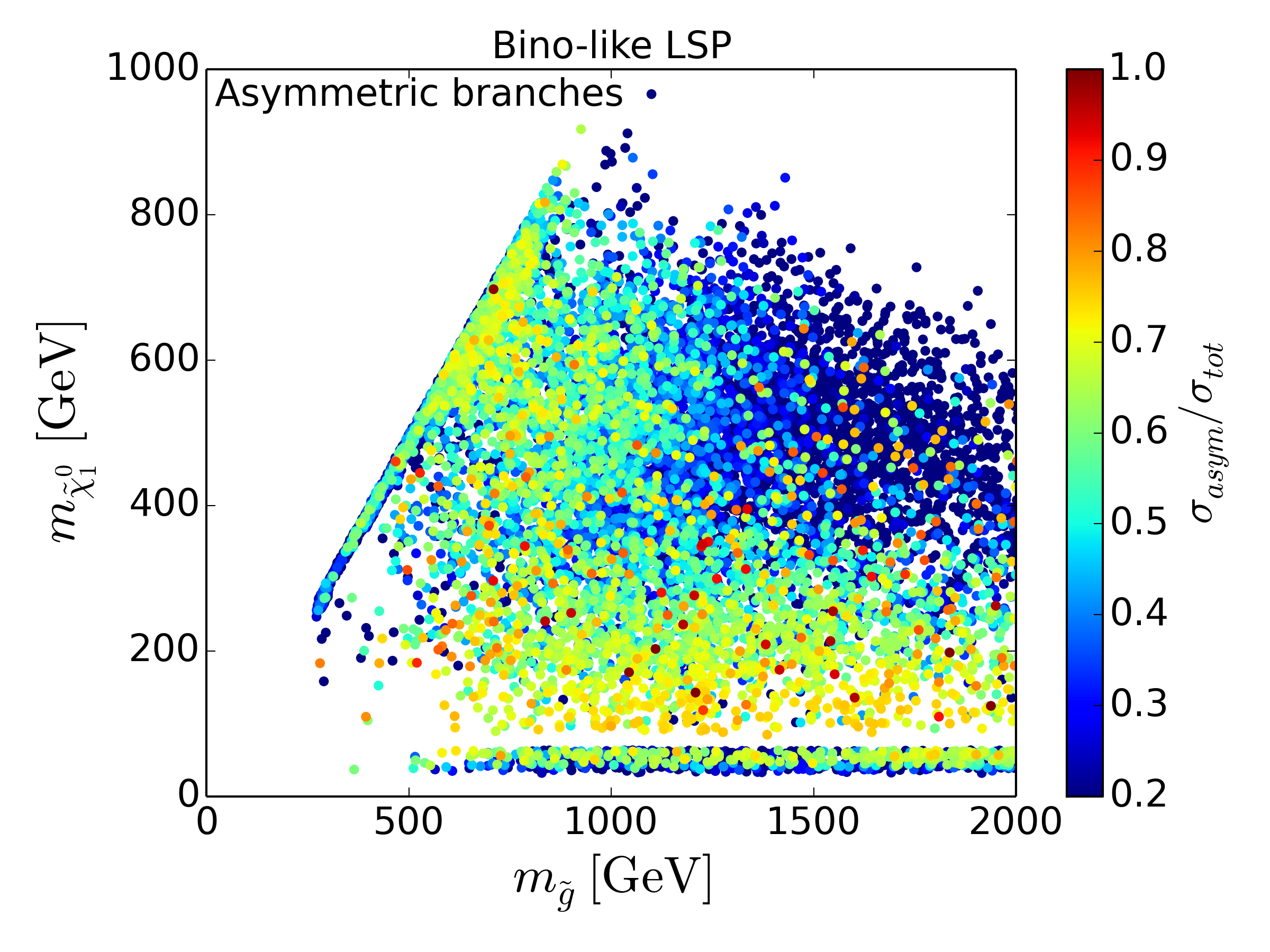}\hfill
\includegraphics[width=0.49\textwidth]{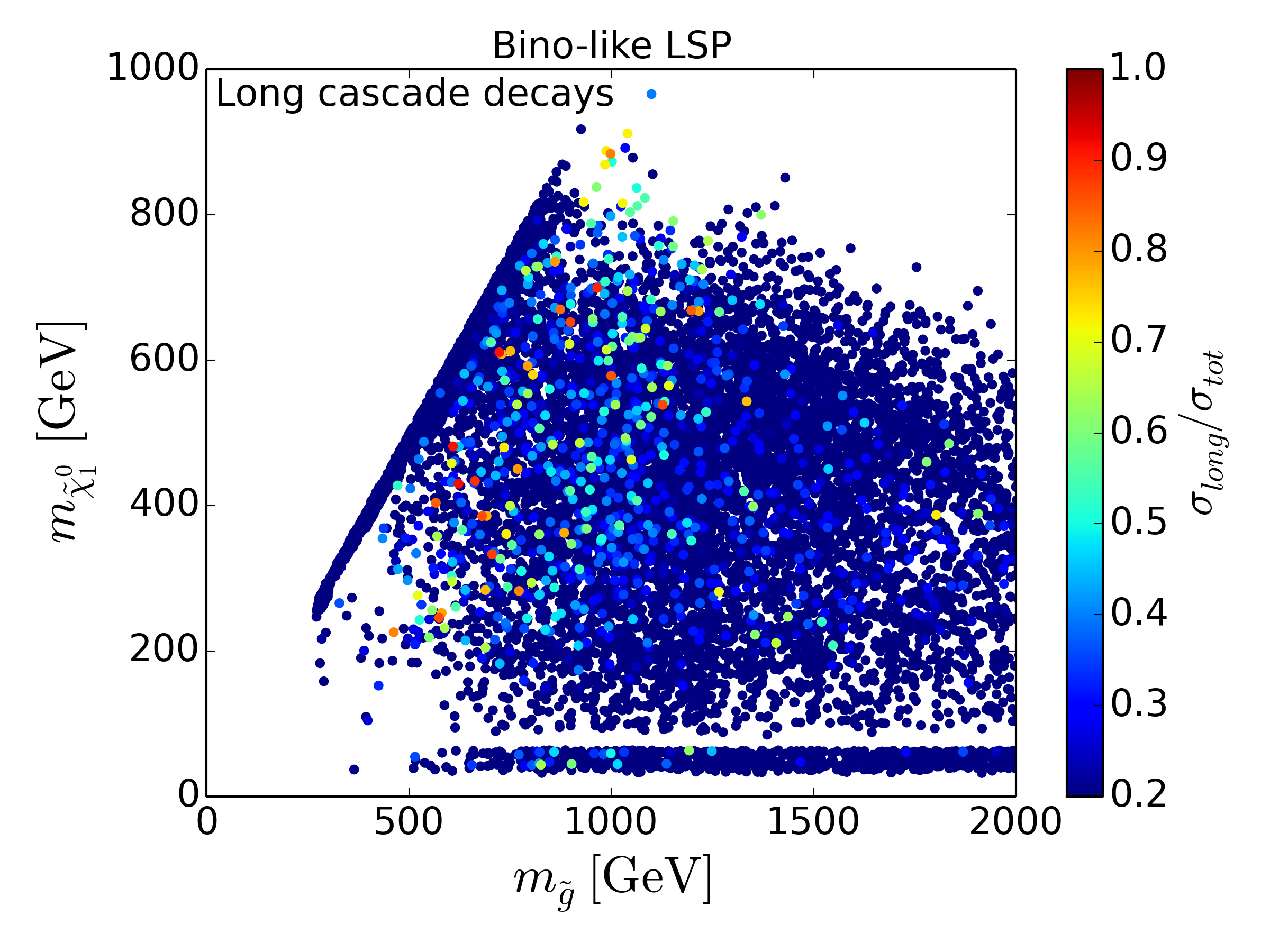}
\caption{Relative cross section in unconstrained decays with asymmetric branches (left) and long cascade decays (right), for scenarios with a
bino-like LSP. Here the total cross section $\sigma_{tot}$ refers to the full $8$~TeV SUSY cross section.
Only allowed points with total cross section larger than $10$~fb have been considered.
\label{fig:pmssmBinoAsym}}
\end{figure}

To understand the possibilities of improving the coverage, without going into details
about the specific missing topologies, we show in Figures~\ref{fig:pmssmBinoAsym}
the relative cross section that goes into missing topologies with asymmetric branches (left) or
long cascade decays (right).
We classify as long cascade decays topologies with two or more intermediate particles,
thus they have at least four free parameters, and we no longer consider a simplified model
description viable.
Short cascade decays are classified as asymmetric branch topologies if
the two decay branches differ, either because the
initially produced particles follow different decay chains, or because the initially produced particles
are different.
We see that in fact missing topologies with asymmetric decay branches are important for a
large number points,
whereas long cascade decay topologies are dominant only in a few points.

Finally we specify that a particularly important missing topology with asymmetric branches arises
from gluino-squark associate production, giving
a 3 jet + $E_T^{\mathrm{miss}}$ final state.
This is important in particular when the light-flavor squarks are highly split and the gluino can
decay to a single on-shell squark.
The relevant process is $pp\to \tilde g\tilde q$ followed by $\tilde q\to q\tilde\chi^0_1$ on one branch
and $\tilde g\to q\tilde q\to q\bar q \tilde\chi^0_1$ on the other branch.
The same topology is also possible when gluinos are lighter than all squarks and decay dominantly via a loop decay to a gluon and the neutralino LSP.
In this case we have $pp\to \tilde g\tilde q$ followed by $\tilde g\to g\tilde\chi^0_1$ on one branch and
$\tilde q\to q\tilde g\to qg \tilde\chi^0_1$ on the other branch.

\section{Conclusions}
In summary we found that in the context of the pMSSM about 55\% (63\%) of bino-like LSP
(higgsino-like LSP) scenarios excluded by ATLAS in a comprehensive event simulation study
can also be excluded using currently available SMS results.
This includes a significant improvement from using EM-type results as well as UL-type results
in the SModelS database.
The coverage is about 10\% lower when using UL-type results only.
Analysing the missing topologies for points escaping all SMS limits, we found that asymmetric decay branches are
by far dominant, while long cascade decays are much less important.
The SModelS EM database should therefore be extended with asymmetric decay branch topologies
to improve the
coverage by SMS results.
A particularly important missing topology is the gluino-squark simplified model.
Note that there are in fact various production and decay channels that are important when both gluinos and squarks are light,
and the relative importance depends the precise mass pattern.
To constrain generic models the efficiencies of each topology should therefore be evaluated separately,
such that the set of efficiency maps can be used to test the total cross section.

\end{document}